# Autonomous epitaxial atomic-layer synthesis via real-time computer vision of electron diffraction

Haotong Liang[1,2,*], Yunlong Sun[3], Ryan Paxson[1,2], Chih-Yu Lee[1,2], Alex T. Hall[1], Zoey Warecki[4], John Cumings[1], Hideomi Koinuma[5], Aaron Gilad Kusne[1,6], Mikk Lippmaa[3,*], and Ichiro Takeuchi[1,2,*]

[1]Department of Materials Science and Engineering, University of Maryland, College Park MD 20742
[2]Maryland Quantum Materials Center, Department of Physics, University of Maryland, College Park MD 20742
[3]Institute for Solid State Physics, University of Tokyo, Kashiwa, Chiba 277-8581, Japan
[4]Materials Science and Technology Division, U.S. Naval Research Laboratory, Washington D.C., USA
[5]Smart Combinatorial Technology, Tokyo 105-0014, Japan
[6]National Institute of Standards and Technology, Gaithersburg, MD
*Corresponding author. Email: hliang16@umd.edu, mlippmaa@issp.u-tokyo.ac.jp, takeuchi@umd.edu

Autonomous science platforms which make decisions on the fly are fundamentally changing the outlook for materials development. AI-driven schemes can effectively reduce the total number of iterations needed to arrive at the best stoichiometry for desired properties or optimum synthesis parameters by significant margins. Here, we demonstrate real-time closed-loop autonomous navigation of a multi-dimensional synthesis parameter space for fabricating phase-pure epitaxial films of a metastable functional oxide phase using pulsed laser deposition. Sequential growth iterations in search of the optimized recipe to stabilize the desired crystal phase were performed using frame-by-frame quantitative computer vision of electron diffraction images at the unit-cell level. Our scheme regularly resulted in > 30-fold reduction in the number of experiments compared to comprehensive parameter-space mapping. The real-time workflow developed here can be readily extended to other thin film synthesis platforms opening the door for self-driving atomic-level materials design as well as autonomous semiconductor manufacturing.

Recent reports of autonomous materials science schemes have shown that Bayesian optimization (BO) techniques can effectively self-navigate and learn a complex multi-dimensional terrain of materials synthesis parameter space, a task that is particularly hard for a human operator to perform optimally(*1–11*). Closed-loop autonomous synthesis workflows can be particularly powerful if the materials characterization that provides structure or property-based feedback at each synthesis iteration can be performed in real time.

For epitaxial thin film growth, various in situ monitoring techniques exist that can provide a powerful "eye" to the growers. RHEED, in particular, can be used to glean complex surface atomic structure information in real time during the growth processes, and it can help to guide the synthesis task(*12*). However, due to the complexity of the epitaxial growth process and the possibility of multi-phase growth, direct quantitative analysis of RHEED images remains challenging, prompting the development of various machine learning-based techniques(*13–17*). In this work, we have developed a computer-vision-based real-time analysis pipeline that derives the required feedback information for Bayesian process optimization from deep-learning image processing of RHEED patterns. The pipeline automates the tracking of RHEED image features arising from simultaneous formation of disparate phases and extracts quantitative lattice parameters of all constituent crystal phases or domains.

The Bayesian process optimization is demonstrated here for pulsed laser deposition (PLD) of oxide thin films. PLD is a complex physical deposition process in which a high-intensity pulsed laser ablates a solid target to generate a transient, high-energy plasma plume that enables stoichiometric transfer and controlled growth of thin films with tailored structural, chemical, or functional properties(*18*). For example, the ambient oxygen pressure during growth primarily controls the valence state of cations in an oxide film, but the film composition, defect density, growth mode, and other properties are also indirectly influenced by the dynamics of the plasma plume where the spatial distribution and the kinetic energy of the ablated ions is affected by collisions with the ambient gas molecules(*19*). The substrate temperature is also a key parameter that not only affect the thermodynamic stability of a particular phase, but also affects the absorption, diffusion, and desorption of the adatoms, indirectly affecting the surface roughness and crystallinity of deposited materials(*20*). The PLD laser repetition rate primarily controls the average growth rate, but it also has a subtle effect on layer nucleation and the extent of surface relaxation between successive deposition pulses, affecting the growth mode and the surface flatness(*21*). Increasing the laser repetition rate can lead to a higher effective substrate temperature due to enhanced energy transfer from the plasma (*22*). These indirect processes influencing the crystal growth are difficult to quantify individually and often ignored by a human operator, whereas a BO model can efficiently optimize the process in an arbitrarily multi-dimensional parameter space while automatically learning the material's growth behavior even in the presence of complex hidden interdependencies between the directly controllable process parameters. This ability makes it easier to add the less-explored process controls to the materials optimization process.

In order to showcase the ability of our self-driving PLD workflow (Fig. 1) to autonomously and rapidly home in on optimized, phase-pure synthesis recipes for a selected compound, we have focused on the growth of hexagonal $TbFeO_3$, which is a representative member of the general h-(RE)$FeO_3$ (RE = rare earth) phases that have the $P6_3cm$ structure and include several known multiferroic compounds with non-collinear antiferromagnetic order(*23*). Unlike the bulk-stable orthorhombic $TbFeO_3$ (Pbnm) counterpart (o-$TbFeO_3$), h-$TbFeO_3$ is a metastable phase that can only be synthesized in the form of nanoparticles stabilized by surface energy(*24*, *25*) or as thin films under epitaxial strain and/or with chemical substitutions (*26*, *27*). High-quality h-$TbFeO_3$ thin films are of particular interest for exploring ferroelectrically controlled spin-torque oscillations with high efficiency, which can play a unique role in antiferromagnetic spintronics with ultrafast terahertz-scale switching(*28*).

## Autonomous Experimental Setup

Fabricating high-quality phase-pure epitaxial h-(RE)$FeO_3$ thin films ordinarily requires an extensive mapping of process parameters, necessitating a large number of systematic deposition experiments coupled with in-depth structural characterization. For the live autonomous synthesis experiment, we have elected to simultaneously optimize three parameters: the oxygen partial pressure, the temperature of the yttria-stabilized zirconia (YSZ) (111) substrate, and the laser repetition rate, covering both thermodynamic and kinetic parameters that govern the formation of materials phases in PLD.

The essential part of the autonomous workflow (Fig. 1) is the frame-by-frame in-situ RHEED analysis of the atomic surface structure evolution during each deposition iteration. We set up the autonomous experiment by defining a quantitative performance measure (PM) that is automatically calculated from the RHEED images and directly quantifies the film characteristics that we wish to optimize. We implemented a Gaussian-process based BO cycle to perform closed-loop optimization of the three-dimensional deposition parameter space in order to maximize the PM. After each experimental iteration of 25 nm thin film growth, the BO suggests a new set of process parameters with which to perform the next experiment that can either reduce the overall uncertainty of the estimated PM landscape or achieve a higher PM value.

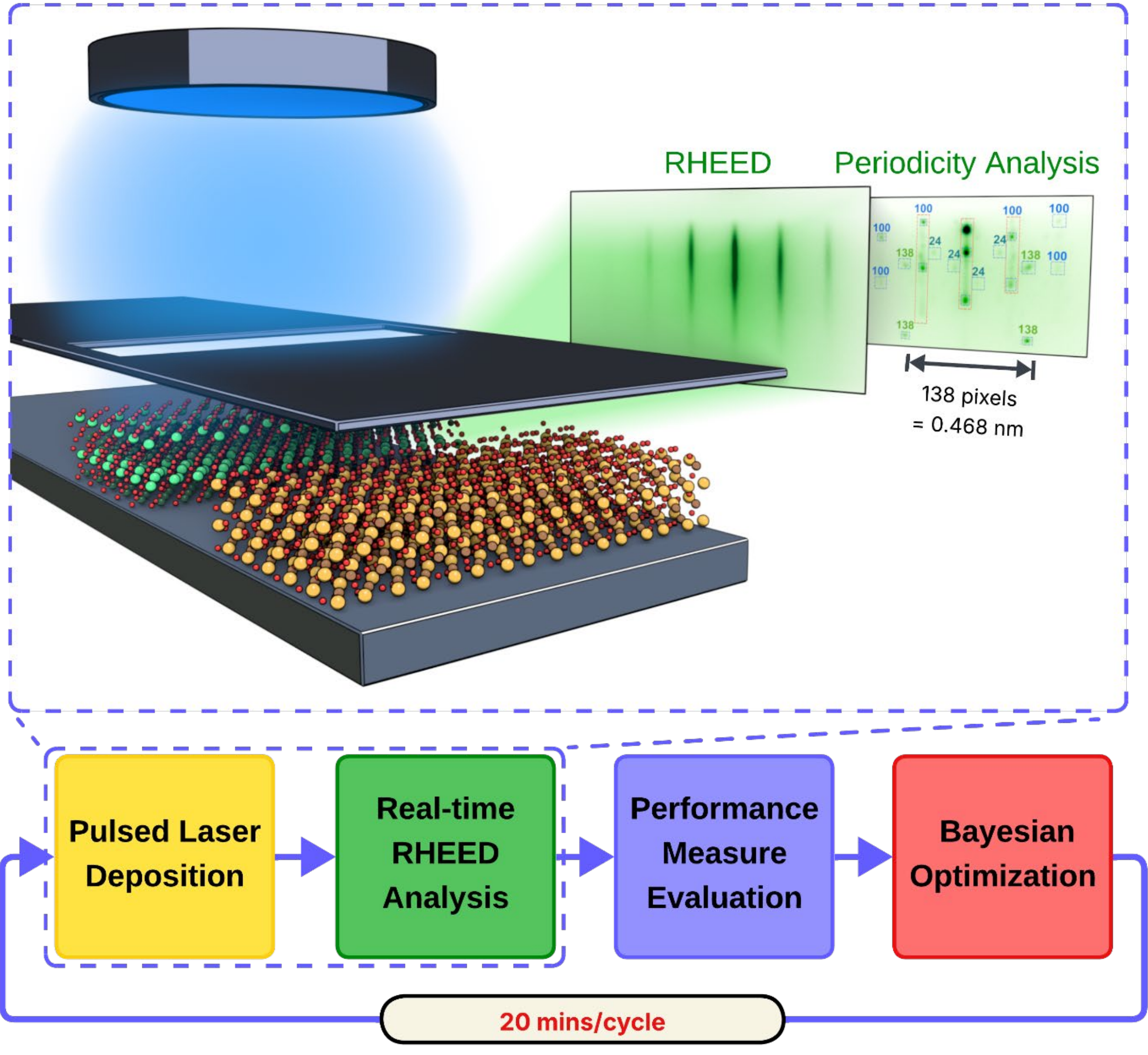

**Fig. 1. Real-time autonomous optimization of epitaxial deposition of oxide thin films**. Multiple films are deposited sequentially by PLD on separated areas on a single substrate using a 0.8 mm wide slit opening on an automated shadow mask (see Supplementary Fig. S1 for details of the sample/substrate configuration). In-situ RHEED and computer vision algorithms are applied to monitor and analyze the growth process during each deposition. After each deposition, based on the growth evolution and the final surface state of the film, Gaussian process based Bayesian optimization decides which deposition parameter values to try next. Each cycle took 20 min to complete for a 25 nm film deposition. The schematic illustrates the second deposition iteration.

The key to an efficient closed-loop thin film synthesis process is the ability to rapidly perform a continuous series of experimental iterations with minimal interruptions. Existing combinatorial PLD systems equipped with computer-controlled shutters can be easily repurposed to sequential spatially-selective depositions on separate regions of a given substrate, as shown in Fig. 1 (and in Supplementary fig. S1), while scanning RHEED is used to monitor the growth dynamics during each deposition. For the h-TbFeO3 growth optimization, each closed-loop cycle took about 20 minutes, the majority of which time was spent on the actual thin film deposition and temperature ramping.

The autonomous synthesis experiment requires automated real-time RHEED image analysis even when some of the phases that occur or the pattern types are not known beforehand, and the live computer vision analysis process has to extract semantic information from the images. We have previously used a U-Net-based(*29*) RHEED-image segmentation method to quantitatively analyze the periodicity of streak patterns within the zero-order Laue circle for phase identification(*30*). Through a systematic growth parameter space mapping study of $Fe_xO$ films on $Al_2O_3$ (0001) substrates, we demonstrated that the RHEED image segmentation method is effective in correctly capturing the abundance ratio of predominant phases, e.g. magnetite ($Fe_3O_4$) and hematite ($Fe_2O_3$), growing simultaneously. For the present work, we used an updated method where the U-Net convolutional neural network (CNN) is replaced with a Cascade Mask Region Based Convolutional Neural Network (Cascade Mask R-CNN) model(*31*), which is an instance segmentation model that produces a separate mask for each diffraction feature type(see Supplementary fig. S2). This allows live quantitative periodicity analysis that yields the corresponding lattice constants as different phases appear and disappear during a growth run. The schematic of the deep learning-based pipeline is shown in Fig. 2A. An example of the model periodicity analysis output based on the segmented instances is shown in Fig. 2B where features with different periodicities measured in image pixels are highlighted. In addition to the previously demonstrated periodicity analysis algorithm, we introduce a tracking algorithm to dynamically monitor the evolution of the periodicity from independent features simultaneously, creating a map of observed periodicities as a function of the deposition time (Fig. 2C). In addition to the instance segmentation task, a classification head is attached to the Cascade Mask R-CNN Resnet backbone to perform classification without a substantial increase in computational cost. This auxiliary classifier is trained to perform binary classification of three main surface features, namely, 2D diffraction (i.e. epitaxial growth), 3D transmission diffraction (i.e. columnar growth), and 3D polycrystalline diffraction, all common features in RHEED(*32*) (Fig. 2C).

Using the tracking algorithm, we monitor changes in the in-plane lattice constant and the stability of phase(s) being grown, as well as the surface morphology as a function of deposition time on-the-fly (Fig. 2C). During the deposition process, the live analysis is carried out at a rate of 2-3 RHEED frames per second. Once the deposition is complete, the algorithm computes the PM that characterizes the growth quality of the h-$TbFeO_3$ film based on three primary objectives: 1) maximize the time window during which phase-pure single-crystal growth of the target phase is taking place; 2) maximize the crystallinity of the target phase, as represented by the FWHM of the diffraction pattern streaks; and 3) reduce the surface roughness by maximizing the decay time of the specular reflection. The three objectives are individually normalized to a 0 to 1 range and then summed together. In addition, we have placed a preferential term in the PM function which aims to speed up experimental time and thus the overall workflow by increasing the laser repetition rate. The final range of the PM is thus 0 to 3.1. A demonstration of the RHEED image analysis pipeline is provided in Supplementary video. S1.

The autonomous experiment commences with the deposition of several h-$TbFeO_3$ thin films at randomly selected points in the 3-dimensional parameter space (substrate temperature 350 °C to 1000 °C, oxygen partial pressure $1.33 \times 10^{-4}$ Pa to 1.33 Pa ($10^{-6}$ Torr to $10^{-2}$ Torr) , and the laser repetition rate 2 Hz to 10 Hz) using the Latin hypercube sampling algorithm(*33*). The PM values derived from the live RHEED image analysis of these initial depositions were used to train a Gaussian Process Regression (GPR) model(*34*), implemented in GPyTorch(*35*), to establish a surrogate model of the PM value landscape in the process parameter space. After the initial training runs, the autonomous run was turned on, and the subsequent sets of process parameters were determined for each iteration using the Upper Confidence Bound (UCB) acquisition function(*36*). After each autonomous-run deposition, the new PM value was evaluated and added to the GPR training data. The autonomous deposition-analysis cycle continues, and the GPR model continues to get updated until the PM converges, i.e. the BO can no longer find a better set of process parameters.

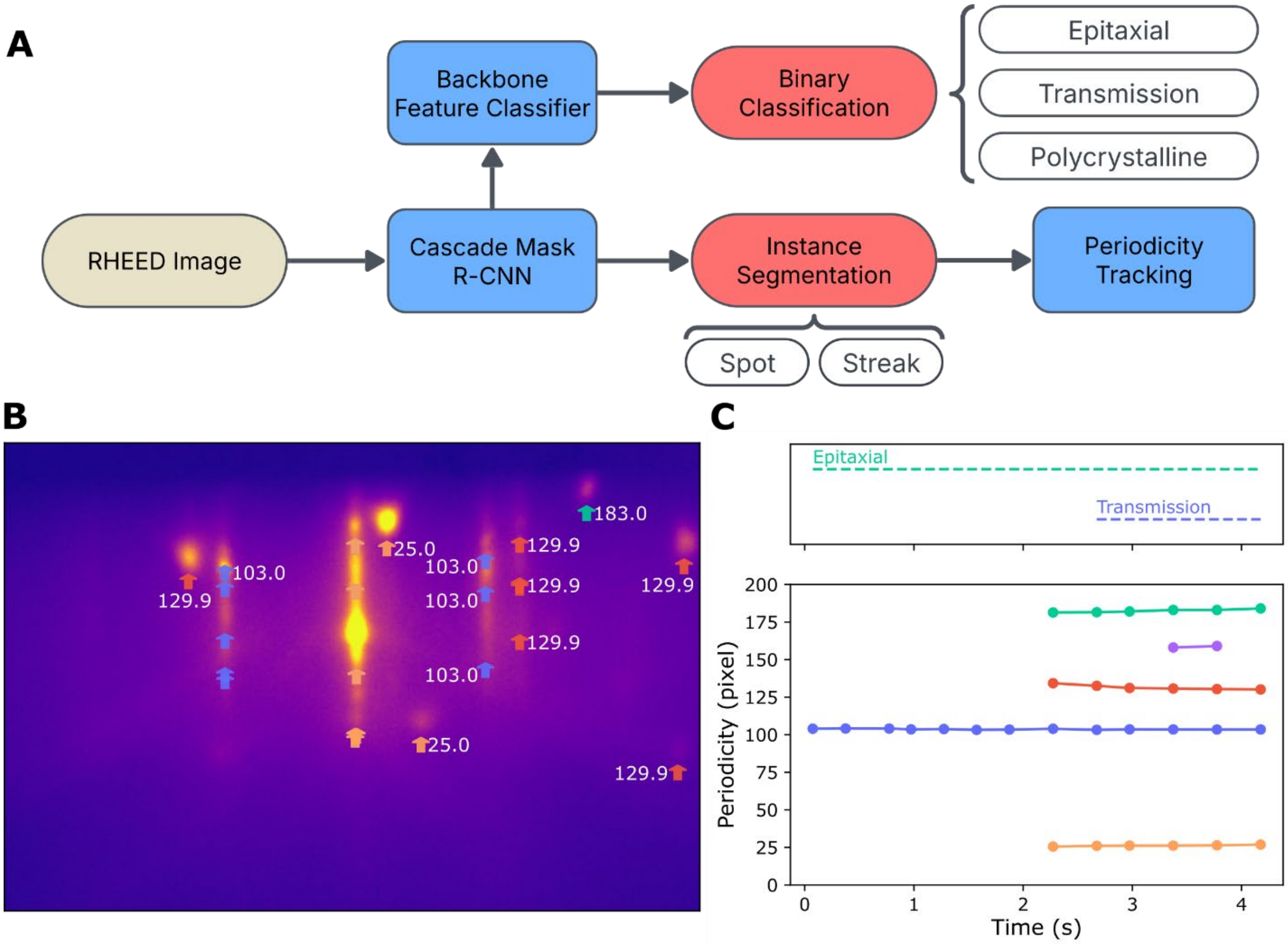

**Fig. 2. Dynamic computer vision for automated analysis of electron diffraction (RHEED) during film synthesis, and detection of secondary phases. (A)** Cascaded Mask R-CNN based RHEED image

analysis pipeline. (**B**) Periodicity (which corresponds to an in-plane lattice constant) measured in image pixels of an example RHEED pattern during the deposition. Periodicity is labeled along with the colored arrows to indicate the location of each segmented feature. (**C**) Periodicity tracking results and binary classification outputs during an example deposition. The periodicity of 103.0 pixels (blue) corresponds to 0.627 nm, the in-plane lattice constant of h-$TbFeO_3$. Appearance of other pixel periodicities suggests the formation of secondary phases indicating that a single-phase film was not obtained in this iteration.

# Results

The detailed optimization pathway of autonomous h-$TbFeO_3$ growth optimization on YSZ (111) substrates is visualized in Fig. 3 by plotting the PM values for each iteration. The RHEED images captured at the end of the depositions are shown in Supplementary fig. S3. In this run, starting with 5 randomly sampled initialization points, the autonomous workflow quickly established a good surrogate model with just 4 additional experimental depositions. The autonomous workflows finally converged at: 1.33 Pa ($10^{-2}$ Torr), 831 °C, and 10 Hz after 27 iterations as the BO gradually shifted from exploration to exploitation (Fig. 3A). The final posterior mean of the GPR model (Fig. 3B) suggests that phase-pure films can be obtained in the temperature window from 700 °C to 900 °C. Within this region, the surface smoothness and crystallinity are improved at higher temperatures. At temperatures above 900 °C, the h-$TbFeO_3$ phase becomes unstable, and a sharp intensity decay is observed within and after the deposition suggesting a marked change in the surface structures. On the other hand, the oxygen partial pressure, up to 1.33 Pa ($10^{-2}$ Torr), monotonically improves the phase stability of the h-$TbFeO_3$ phase, allowing it to be deposited at elevated temperatures for better film quality. Given the number of all possible sets of values in the three-deposition parameter space (≈ 10 x 10 x 10), being able to arrive at the optimum point in less than 30 iterations corresponds to a roughly 30-fold reduction in the required number of experiments compared to a comprehensive grid search.

After the autonomous experiment, a single 25 nm h-$TbFeO_3$ film was deposited on a 10 mm × 10 mm x 0.5 mm YSZ (111) substrate using the identified conditions (1.33 Pa; 831 °C; 10 Hz) to validate the optimization result. Following the deposition, the RHEED patterns were recorded at different azimuthal angles by rotating the substrate in plane ($\phi$ = 0 °, -30 °, and 30 °). The periodicity measured at $\phi$ = 0 ° and 30 ° was found to exhibit a $1{:}\sqrt{3}$ relationship while those between $\phi$ = −30 ° and 30 ° have 1 : 1 relationship indicating that the surface structure had the correct 6-fold symmetry (Fig. 4A). The in-plane lattice parameter was estimated to be 6.2 Å from RHEED. Details of the RHEED images and periodicity evaluation are shown in Supplementary fig. S4. In fig. 4b, the out-of-plane $2\theta/\omega$ X-ray diffraction (XRD) shows only h-$TbFeO_3$ (002), (006), and (008) peaks, yielding a c-axis lattice parameter 11.87 Å. The $\varphi$-scan of the h-$TbFeO_3$ (112) peaks in Fig. 4C shows the expected 6-fold symmetry. Comparison with the YSZ substrate (220) peaks shows that the film-substrate epitaxial relationship is h-$TbFeO_3$ (110) // YSZ (110) and h-$TbFeO_3$ (001) // YSZ (001). Thus, in-situ RHEED and XRD results indicate that the deposition conditions are indeed optimized for achieving phase-pure h-$TbFeO_3$ films. An atomic force microscopy (AFM) image (Fig. 4D) of the as-deposited surface shows large terraces with single unit cell steps of about 1.0 nm, indicating that the roughness measure component in the PM was a good surrogate for the actual surface roughness. A scanning transmission electron microscopy (STEM) image captured at the interface between the $TbFeO_3$ film and the YSZ substrate shows alternating Tb-O and Fe-O layers, proving that the film is indeed h-$TbFeO_3$ (Fig. 4E). The non-centrosymmetric out-of-plane displacement of the Tb atoms agrees with a simulated image along the h-$TbFeO_3$ $[0\bar{1}0]$ zone axis, suggesting that the h-$TbFeO_3$ film adopts the multiferroic structure of space group $P6_3cm$.

A thicker, 75 nm film was deposited to further examine the stability window of the h-$TbFeO_3$ and study the magnetic properties of the material. An out-of-plane field warming measurement (Fig. 4F) shows a peak in the magnetic moment with a local maximum at 37 K which agrees with a report of an antiferromagnetic – weak ferromagnetic (AFM-wFM) transition temperature at 30 K(*37*). The weak-ferromagnetic component decreases with increasing temperature and vanishes at the Néel temperature of 119 K, close to the reported Néel temperature range of 110 K to 120 K(*27*, *37*)

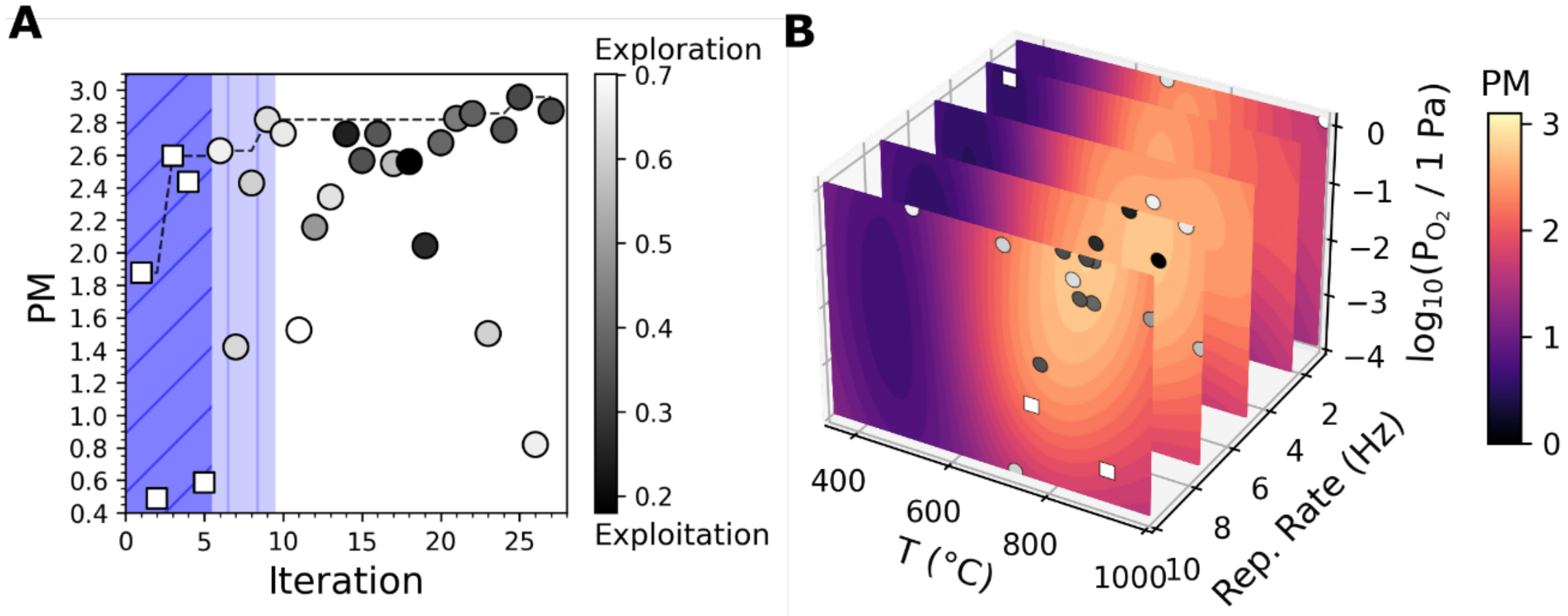

**Fig. 3. Optimization pathway of the autonomous growth of h-$TbFeO_3$.** (**A**) Obtained performance measure (PM) value at each iteration of the deposition. The first five depositions, selected by Latin hypercube sampling algorithm, are shown as squares within the dark blue hatched region. A good surrogate model was established after visiting points in the light blue hatched region. The dotted line shows the highest PM value obtained up to that iteration, and it tracks the progress of optimization. (**B**) Contour plots of the GPR posterior mean of the PM value at the last iteration in the autonomous experiment. Each contour plot is a landscape of oxygen partial pressure $P_{O_2}$ and substrate temperature T slice at a particular laser repetition rate. The color of the scatter points in (A) and (B) indicates the contribution of the exploration term $\beta\sigma(x)/\big(\beta\sigma(x) + \mu(x)\big)$ in the acquisition function, where $x$ is a three-dimensional deposition parameter vector: brighter points correspond to higher weight for exploration whereas darker points correspond to higher weight towards exploitation. The cluster of points near the optimum indicates the convergence of the GPR model.

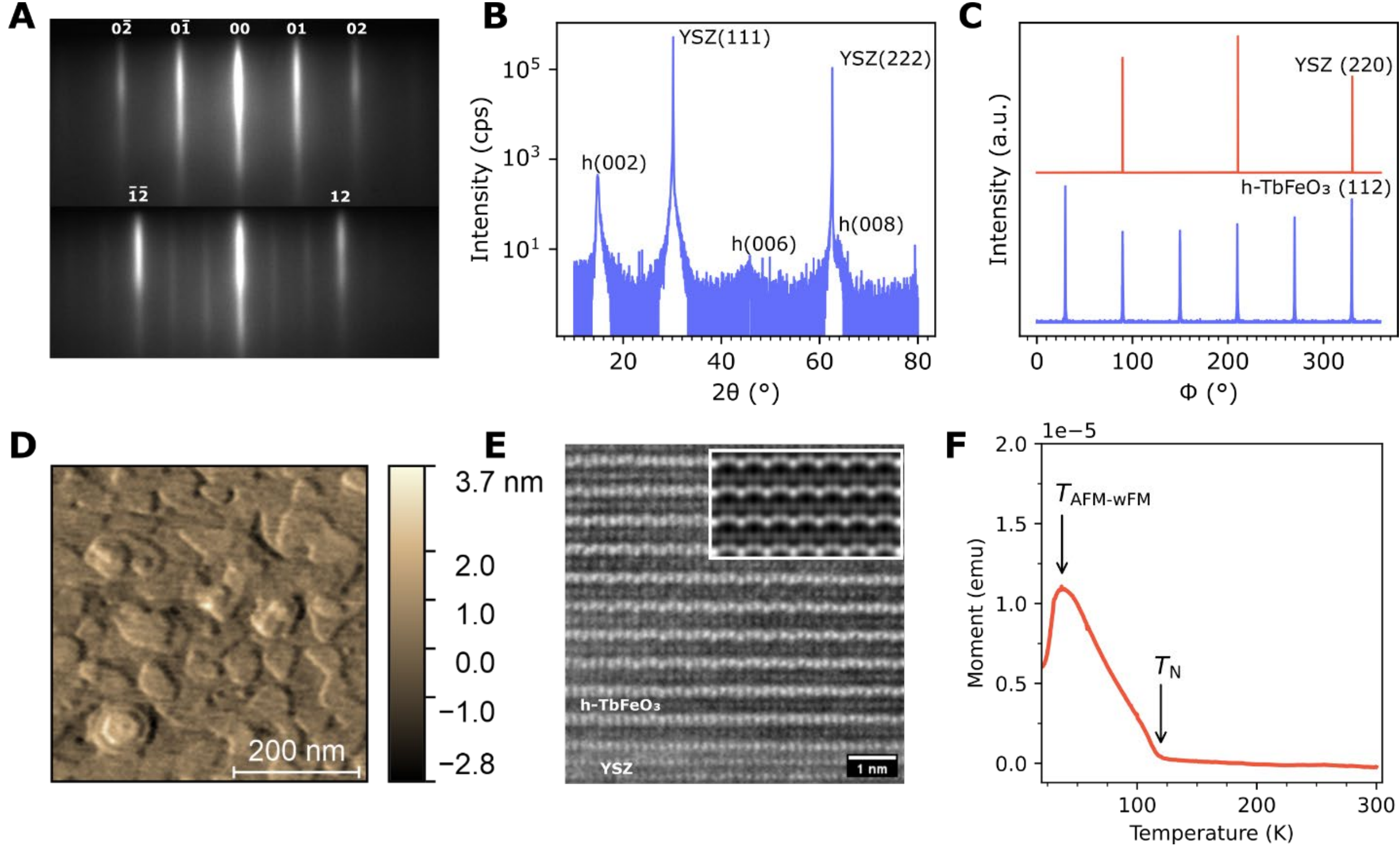

**Fig. 4: Optimized phase-pure growth of h-$TbFeO_3$ and its properties.** (**A**) RHEED images taken at azimuth angle $\phi = 0$ ° (top) and -30 ° (bottom). (**B**) Out-of-plane XRD measurement of optimized $TbFeO_3$. h-$TbFeO_3$ peaks are marked with h. (**C**) $\phi$-scan of h-$TbFeO_3$ (112) peak vs YSZ (220) peak. (**D**) AFM image of the h-$TbFeO_3$. Height difference between terraces is 1 nm or one unit cell of h-$TbFeO_3$. (**E**) STEM image taken at the interface between YSZ (111) substrate and h-$TbFeO_3$. The zone axis is h-$TbFeO_3$ $[0\bar{1}0]$. The inset shows a simulated STEM pattern based on the h-$TbFeO_3$ structure (space group $P6_3cm$). (**F**) Magnetization vs temperature curve at $7.96 \times 10^4$ A/m (1000 Oe) in the YSZ (111) direction. A clear AFM-wFM transition is observed at 37 K while the Néel temperature is 119 K.

We have carried out further validation experiments to understand the identified three-dimensional PM landscape by depositing films at selected conditions near, but not at the peak of the PM. (The details are in the Supplementary.) These experiments revealed that the identified conditions are the result of competition between crystallinity, which requires elevated growth temperatures above 700 °C, and metastable phase stability, which favors temperatures below 900 °C and high oxygen pressure. The preference for high laser repetition rate is due to the suppression of the h-$TbFeO_3$ to o-$TbFeO_3$ transition as it can limit the time that the film is exposed to high temperatures.

We have performed the same autonomous thin film synthesis for several other h-(RE)$FeO_3$ phases, namely, h-$GdFeO_3$ and h-$EuFeO_3$, and obtained similar trends in the optimum deposition parameters as the one shown in Fig. 3B, with the optimum substrate temperature centered around 850 °C. The deposition conditions and the corresponding PM values for h-$TbFeO_3$, h-$GdFeO_3$, and h-$EuFeO_3$ are listed in the supplementary table S1 – S3. In the literature, reported PLD deposition conditions for successful h-(RE)$FeO_3$ thin films are in the similar range, ≈1.33 Pa (10$^{-}$

$^{2}$ Torr) to 13.3 Pa ($10^{-1}$ Torr), 750 °C to 900 °C, but are generally scattered in parameter values and do not convey a converging behavior as captured in the present work(*27*, *37–50*), indicating that prior to the present work, the parameter space was largely underexplored. More information regarding the PM landscape is available in the supplementary materials.

## Discussion and outlook

We have demonstrated a fully autonomous thin film optimization workflow incorporating live quantitative analysis of in situ RHEED images taken during deposition. Through deep learning and segmentation, we were able to accurately decipher on the fly the rich atomic-scale surface structure information from multiple competing growth mechanisms, allowing us to define a robust film quality PM needed to effectively drive the self-navigating scheme, which led to a 30-fold reduction in the number of experimental iterations compared to mapping the entire parameter space.

Because local crystal structure and defect formation are direct factors governing the properties of functional thin film materials at large, the present workflow can be expanded to a variety of other deposition schemes and materials systems. Because our dynamic deep-learning and segmentation analysis allows robust periodicity and, in turn, in-plane lattice constant determination, it can be applied to *any* materials including new or hitherto unknown phases, and it is impervious to extraneous effects such as change in the RHEED screen brightness due to the substrate temperature change or shift in the center beam position, which can often undermine more conventional RHEED analysis modes.

The current workflow represents a nimble and agile approach to autonomous systems in that it is entirely based on python-based control and analysis applied to semi-standard existing automated hardware, and it can be applied to any chambers with similar RHEED and sequential deposition capabilities, without the need for more elaborate robotic-based tools. The method can be extended to any number of other process parameters beyond the three we have varied here as long as computerized control is available(*5*, *15*).

The scope of PM can be dictated by the user and depends on the specific goal of the autonomous experiment. The GPR produces a multi-dimensional parameter space map (Fig. 2B) that can be used as a blueprint to further tune the deposition recipe within the defined bounds of the parameter space contour, as often practiced in the fabrication of epitaxial complex multi-component oxide thin films with competing/conflicting deposition requirements, e.g. kinetics vs. stoichiometric or crystallinity control.

While it is straightforward to extend the current work using the combinatorial PLD to autonomous composition mapping campaigns in search of composition(s) with particular structures (or a structural transition, e.g. in a morphotropic phase boundary piezoelectric) for materials discovery, we envision that a more widespread utility of the autonomous approach might be in semiconductor manufacturing settings, where there is a ubiquitous and perpetual need for process optimization and monitoring.

# Methods

## Experimental Setup

Thin film deposition was carried out using a combinatorial PLD system from Pascal Co (Osaka Japan) with a scanning high-pressure RHEED attachment for in-situ characterization. $TbFeO_3$ ceramic targets from Toshima Manufacturing Co had a purity of 99.9%. The base pressure of the deposition chamber was $8 \times 10^{-7}$ Pa ($6 \times 10^{-9}$ Torr). The autonomous experiment's process parameter search range was $1.33 \times 10^{-4}$ Pa ($10^{-6}$ Torr) to 1.33 Pa ($10^{-2}$ Torr) for oxygen partial pressure, 350 °C to 1000 °C for substrate temperature, and 1 Hz to 10 Hz for the laser pulse repetition rate. The ablation was done with a Compex Pro 110F KrF 248 nm excimer laser and a pulse energy of 88 mJ. The fluence on the target was 2 J/cm$^2$ in all depositions. For each autonomous deposition, 1000 laser shots were fired at the $TbFeO_3$ target while raster scanning. The total deposition time varied from 100 s to 1000 s, depending on the selected laser repetition rate. The temperature ramp rate was 30 °C/min. XRD measurements were carried out using the Rigaku SmartLab X-ray diffraction system in parallel-beam setting with a Ge (220) monochromator, 5° receiving Soller slit, Cu x-ray source, and a HyPix-400 2D detector. The AFM image was acquired in tapping mode on a Digital Instrument 5000 microscope using a Nanoworld ARROW-NC-SPL probe. The resonance frequency of the probe was 285 kHz and imaging was conducted at a drive frequency of 284.8 kHz. Cross-sectional lamellae for STEM imaging were prepared using a ThermoFisher Helios G3 DualBeam Focused ion beam scanning electron microscope, with final argon ion-polishing at 900 eV in a Fischione 1040 Nanomill. High-resolution STEM imaging was performed using a JEOL NEOARM operated at 200 kV with a CEOS ASCOR probe corrector. STEM images were obtained by summing 10 to 20 images with dwell times of 0.8 s to1.6 s with a resolution of 1024 x 1024 pixels. The effects of specimen drift were corrected in the resulting summed images. The magnetization measurements were done in a Quantum Design MPMS-XL System. The temperature dependent measurements were done at $7.96 \times 10^4$ A/m (1000 Oe).

## Instance Segmentation Model

The key component of the computer vision analysis pipeline is the feature recognition algorithm, which is implemented via the Cascade-MaskRCNN instance segmentation model(*31*). The Cascade-MaskRCNN model improved upon the real-time object detection model, Faster-RCNN(*51*), by replacing the detection head with a cascaded detection scheme. The cascaded detection scheme works by iteratively refining bounding box prediction quality using the bounding box prediction from previous layers of the model. The adaptation of the cascaded detection scheme increases the bounding box hypothesis quality (i.e., the produced box that has a high intersection over union (IoU) against the ground-truth box in the label) while maintaining its performance at different ranges of IoU. Similar to the original Faster-RCNN algorithm, the Cascade-MaskRCNN model uses a ResNet-based backbone(*52*), and the model used in this work is pre-trained on the COCO dataset and is available through Openmmlab(*53*). The model is fine-tuned on our custom dataset with 3100 annotated RHEED images to output instance segmentation masks for each designed feature. Here, diffraction spot, diffraction streak, and direct beam location are the three target features that the model is trained to recognize. A typical output of the model is shown in supplementary Fig. S4.

### Classification Model

In addition to the instance segmentation task, the Cascade-MaskRCNN could be extended to perform classification without a substantial increase in computational cost. We attached a classification multilayer perceptron model at the tail of the backbone ResNet model. Thus, the features used to perform the instance segmentation are shared with the binary classification model. This auxiliary classification model is trained on top of the trained main instance segmentation mentioned in the previous section, and the backbone ResNet model's weights are frozen during the training. The dataset for the classification uses the same set of RHEED images for the instance segmentation model and is constructed via hand-labeling images into three categories: polycrystalline, epitaxial (2D layer by layer growth), and transmission (3D island growth). The idea behind these categories is to rapidly determine the state of the deposition in a data-driven fashion such that further extension of the capability could be implemented at a similar cost to the original effort.

### Tracking

Time dynamics are a critical aspect of PLD experiments since they not only reveal how materials grow in real-time but also capture the stability of the target crystalline structure during and after the deposition. Autonomously monitoring these dynamics therefore requires an algorithm that can track temporal changes in phase composition and crystallinity. Phase identification was first demonstrated in the RHEED phase mapping experiment(*30*), where periodicities of streak features were used to determine the associated phases. A light-weight periodicity analysis algorithm was developed to iteratively labels spot, or streak features that share the same periodicity into a same phase. This method was used to identify the occurrence of different epitaxial phases with distinct lattice parameters. Further development of this algorithm in this work allows periodicity to be tracked over time. The working principle of periodicity tracking is similar to a typical multi-object tracking algorithm (e.g. IoU tracking(*54*)), where a two-step approach is used. The rough pseudocode is provided in supplementary materials. The periodicity tracker can easily visualize certain phase appearances and disappearance, which provides vital information for process parameters optimization. For each tracked periodicity, further analysis of its evolution can reveal how the lattice constant changes as a function of film thickness, temperature, pressure, and other growth parameters.

## Measures

To optimize the growth of the epitaxial phase, we proposed a phase measure and a quality measure that quantify how long the growth of the target phase is sustained during the deposition and the crystallinity of the final product. The phase measure $m_p$ is derived from the periodicity tracking analysis and binary classification result as shown in Equation 1:

$$m_p = \sum_{t \in \boldsymbol{t}, \chi \in \boldsymbol{\chi}(t)} \frac{\Delta t}{t_s} \quad (1)$$

where $\chi$ is the target phase with $\chi(t)$ being the set of phases that appear at time $t$ which is a discrete time step in the set $\{t_0, t_1, \dots, t_f\}$. The crystallinity of the final product is evaluated by the quality measure $m_c$

$$m_c = 1 - \frac{min(K, \mu_w)}{K} \quad (2)$$

where $\mu_w$ is the average streak width in pixels of the target phase and $K$ is a user-defined factor to ensure that the metric remains positive. Since the evolution of the film roughness is directly correlated with the specular signal, we defined the following roughness measure to quantify the decay of surface quality over time:

$$m_r = \sum_{t \in \boldsymbol{t}} \frac{I(t)\Delta t}{I(t_0)(t_f - t_0)} \quad (3)$$

where $\mathrm{I(t)}$ is the average RHEED intensity within a user-defined bounding box that covers the majority of the specular reflection at the time step t. The user-defined bounding box is kept unchanged throughout the experiment. Additionally, a preferential term, $m_s$ (i.e., a speed measure), is introduced solely to reduce growth time:

$$m_s = \frac{r}{r_m} \quad (4)$$

where $r$ is the ablation laser pulse rate and $r_m$ is the maximum value of the ablation laser pulse rate. Since a fixed number of pulses is used for each run in this experimental setup, a higher pulse rate results in a shorter deposition time. The performance measure (PM) used for the autonomous experiment is the weighted sum of all four mentioned measures:

$$M = w_p m_p + w_c m_c + w_r m_r + w_s m_s \quad (5)$$

Where $w_p$, $w_c$, $w_r$, $w_s$ are user-defined weight for the phase, crystallinity, roughness and speed measures, respectively. In this study, $w_p = 1$, $w_c = 1$, $w_r = 1$, and $w_s = 0.1$ to balance the contributions of different components.

## Bayesian Optimization

At the core of the growth optimization process is the Bayesian Optimization (BO) algorithm(*55*). Here, a type of non-parametric model called Gaussian Process(*34*) is used as a surrogate to learn the underlying parameter space. The autonomous experiment proceeds in an iterative fashion where the GPR model proposed a set of parameters that maximize the acquisition function. A deposition experiment is performed at those parameters, and the Cascade-MaskRCNN model-based pipeline is deployed to calculate the PM value. The experimental conditions and the corresponding PM then would be recorded into the database, and the GPR model is retrained using the updated database.

Considering the complexity of the parameter space associated with PLD, RBF kernels are used for each process parameter (growth temperature, $O_2$ partial pressure, and excimer laser repetition rate). The formulation of the RBF is shown below:

$$K(x_1, x_2) = \sigma^2 \cdot \exp\left(\frac{-||x_1-x_2||^2}{2l^2}\right) \quad (6)$$

where $x_1$, $x_2$ are two input scaler parameters, $\sigma^2$ is variance and $l$ is length scale of the GPR model. Here, a separate set of variance and length scales is used for each process parameter.

At each iteration, the GPR model is trained to predict the mean of the PM values given process parameters and to estimate uncertainty as its covariance of the prediction. We use the UCB acquisition function to balance the exploration and exploitation of BO process. The UCB acquisition function $\alpha(x)$ is defined as follows:

$$\alpha(x) = \mu(x) + \beta \cdot \sigma(x) \quad (7)$$

where $x$ are the input parameters, $\beta$ are user-defined terms to balance the exploitation and exploration, $\mu$(x) and $\sigma(c)$ are the GPR's mean function and the square root of the covariance function, respectively. The next process parameter is then set to be $x^*$ that maximizes the acquisition function:

$$x^* = \arg\max_x a(x) \quad (8)$$

Once the parameter $x^*$ is determined, the next iteration of BO process proceeds until the best PM value converges. The model is considered converged when two or more datapoints are evaluated near the current maximum and shows no more than 5% deviation in the target measure from the maximum. In this study, we found $\beta = 3$ could encourage exploration which is critical for avoiding premature convergence to a local maximum. The hyperparameter tunning is done on a simulated 3-dimentional landscape with two multinomial gaussian peaks.

# Acknowledgement

**Funding:** The development of the autonomous workflow, thin film deposition, and characterization was supported by 3DFeM2, an EFRC funded by the U.S. DOE, Office of Science, Basic Energy Sciences under Award Number DE-SC0021118. The equipment was acquired with the support by the Office of Naval Research grant N000142212455 through DURIP. QuantEmX grant from ICAM and the Gordon and Betty Moore Foundation through Grant GBMF9616 also supported the project. The work at the University of Tokyo was supported by Japan Science and Technology Agency (JST) (Grant No. JPMJMI21G2) and JSPS KAKENHI (Grant No. 25K01650).

**Disclaimer:** All data needed to evaluate the conclusions in the paper are present in the paper and/or the Supplementary Materials. Disclaimer: Certain equipment, instruments, software, or materials are identified in this paper in order to specify the experimental procedure adequately. Such identification is not intended to imply recommendation or endorsement of any product or service by NIST, nor is it intended to imply that the materials or equipment identified are necessarily the best available for the purpose. These opinions, recommendations, findings, and conclusions do not necessarily reflect the views or policies of NIST or the United States Government.

**Correspondence and requests for materials** should be addressed to Haotong Liang, Mikk Lippmaa, or Ichiro Takeuchi.

**Author contributions:** I.T. and M.L. supervised the project. I.T, M.L., H.L., and A.G.K. conceived the experiments. H.L. developed the algorithms. H.L., Y.S., R.P., and M.L. did the growth experiments. A.T.H, Z.W., and J.C. performed electron microscopy. C.L. carried out magnetic characterization. M.L. and H.K. developed combinatorial pulsed laser deposition. H.L., I.T., and M.L. wrote the manuscript. All authors commented on the manuscript before its submission.

**Competing interests:** The authors declare no competing interests.

# Supplementary Materials for

Autonomous epitaxial atomic-layer synthesis via real-time computer vision of electron diffraction

Haotong Liang et al.

Corresponding author: Ichiro Takeuchi, Email: takeuchi@umd.edu

This PDF file includes:

Supplementary Text

Figs. S1 to S6

Tables S1 to S4

References

Other Supplementary Materials for this manuscript include the following:

Movies S1

**Supplementary Text**

## Investigation of the PM Landscape

The PM landscape in the deposition parameter space determined by the present autonomous workflow was further investigated through additional thin-film depositions (25 nm thick) at three conditions near the optimal point (a: 1Hz, 831 °C, 1.33 Pa; c: 1 Hz, 1000 °C, 1.33 Pa; d: 10 Hz, 1000 °C, 1.33 Pa) and at the optimum condition (b: 10 Hz, 831 °C, 1.33 Pa). Immediately following the deposition, each sample was held at the same conditions for 5 minutes to examine the stability of the h-$TbFeO_3$ phase. fig. S5E to fig. S5H shows the time-dependent specular reflection intensity observed during and after the growth for the four samples, starting from when the deposition commenced. Both samples c and d show a significant decay in the specular reflection intensity after the deposition whereas the specular intensity of samples a and b remain stable. In addition, sample c has a more pronounced decay in the intensity during the deposition than sample d indicating that excessively high temperature may reduce the phase stability of the h-$TbFeO_3$ phase and this effect is cumulative in time. XRD patterns of samples a, b, c, and d are shown in fig. S5A to fig. S5D. The orthorhombic $TbFeO_3$ (o-$TbFeO_3$) peaks are observed in samples c and d indicating that the decay of the specular reflection intensity can be associated with a transition from h-$TbFeO_3$ to o-$TbFeO_3$. Such a transition at temperature >900 °C has been observed in a growth condition study of $YbFeO_3$(*49*) and is accurately captured by the autonomously resolved PM landscape. Thus, the discovered optimal conditions are the result of competition between crystallinity which requires elevated temperatures above 700 °C and metastable phase stability which favor temperatures below 900 °C and oxygen pressure ≥ $1.33\times10^{-2}$ Pa ($10^{-4}$ Torr). The preference for higher laser repetition rate is due to the suppression of the h-$TbFeO_3$ to o-$TbFeO_3$ transition as it can limit the extended time exposure of the sample to high temperatures.

To better understand the observed condition trend, we have surveyed the optimal growth conditions and their lattice parameters of h-(RE)$FeO_3$ (RE = Eu, Gd, Tb, Dy, Ho, Er, Tm, Yb, and Lu) (*27*, *37–50*). The most common substrate temperature is 850 °C which is close to the 831 °C identified in this work. Despite the fact that a common oxygen partial pressure is in ≈13.3 Pa ($10^{-1}$ Torr) range, there are few reports which demonstrate the stabilization of the hexagonal structure in ≈1.33 Pa ($10^{-2}$ Torr) range suggesting that the process parameter space is still largely underexplored. The in-plane and out-of-plane lattice parameters for stabilized h-(RE)$FeO_3$, grouped by the substrate materials, are shown in fig. S6 and table. S4. The in-plane lattice parameter of h-$TbFeO_3$ measured in this work follows the apparent trend of increasing in-plane lattice constant with decreasing rare-earth atomic number. The reported c-plane has no generic trend across rare earth elements but is reported to be affected by the film thickness, oxygen partial pressure, and substrate material. The measured out-of-plane lattice parameter is within the range of reported values.

## Pseudocode for RHEED Analysis Pipeline

**Data**: RHEED video stream

**Result**: Final Periodicity, plus a record of Periodicity at each frame

Periodicity ← {}

PeriodicityLog ← [] // Empty list to store periodicity after each frame

**For each** *frame in RHEED video stream* **do**

  PreviousPeriodicity ← Periodicity

  Segmentations ← CascadeMaskRCNN(frame)

  Distances ← GetHorizontalDistanceFromCenter(Segmentations)

  (UnmatchedDistances, Periodicity) ← MatchWithKnownPeriodicity(Distances, PreviousPeriodicity)

  NewPeriodicity ← PeriodicityAnalysis(UnmatchedDistances)

  Periodicity ← Combine(Periodicity, NewPeriodicity)

  // Record the current Periodicity in our log

  PeriodicityLog ← Append(PeriodicityLog, Periodicity)

**End for**

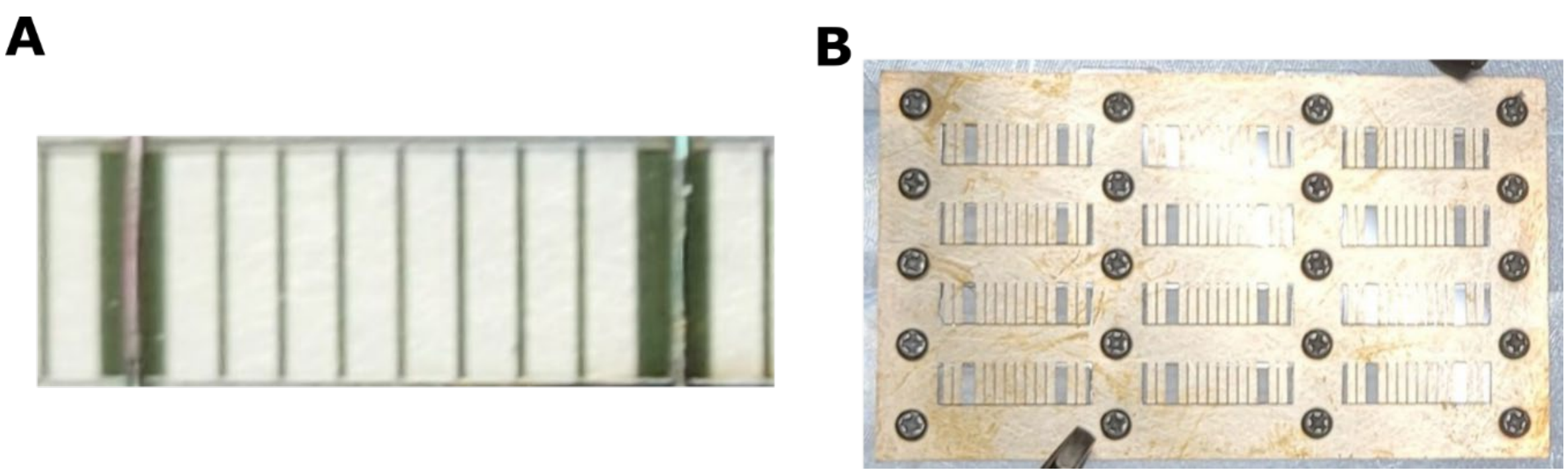

**Fig. S1. Experimental setup for the multi-position deposition.** (**A**) A single (15 x 5 x 0.5) mm YSZ (111) substrate with a thin-layer of gold serving as a fiducial marker. Thin film samples are deposited sequentially within the uncoated regions along the longitudinal direction. (**B**) Shadow mask used to mass-produce the gold markers on YSZ (111) substrates.

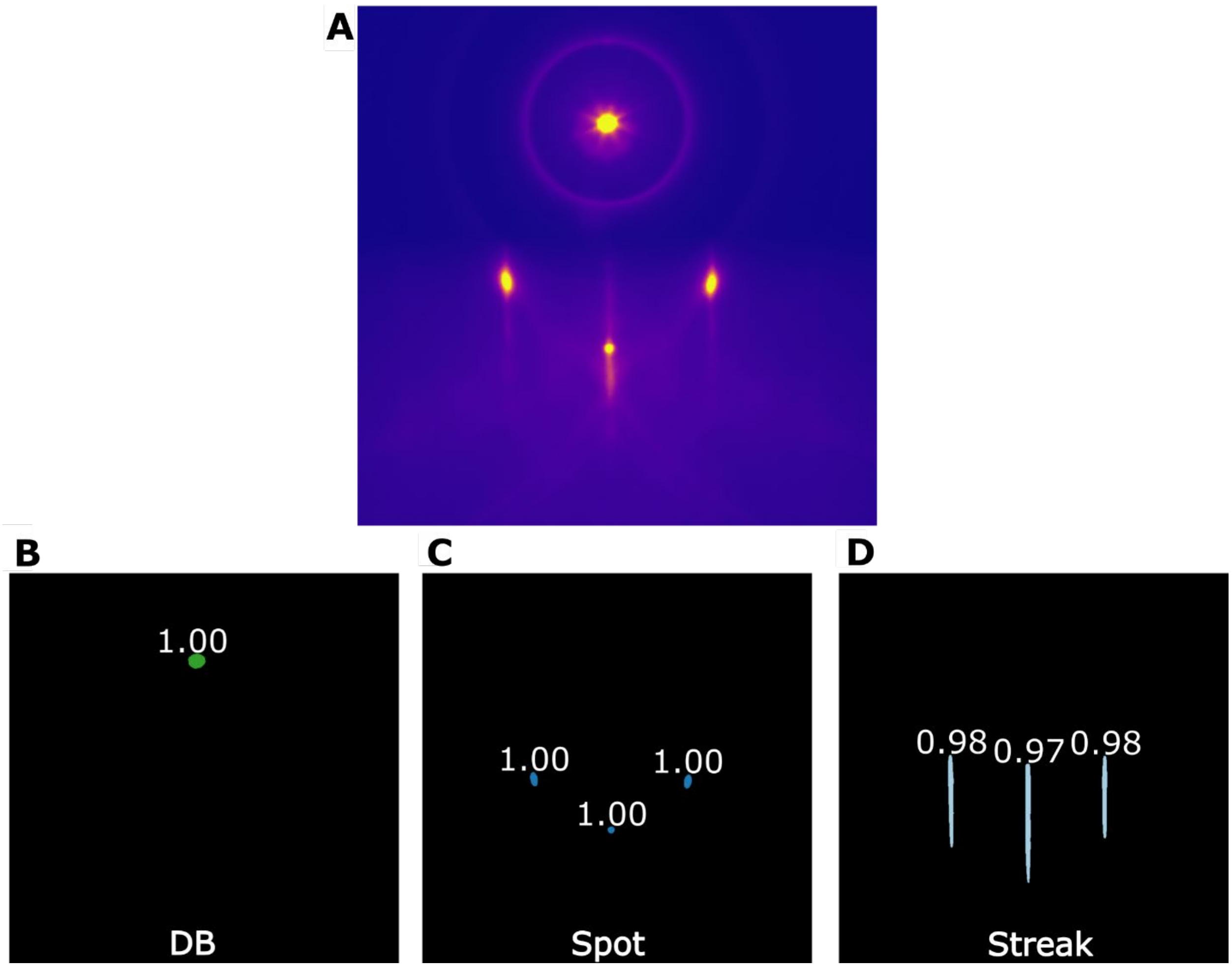

**Fig. S2. An example output of the Cascade-MaskRCNN model given a RHEED pattern of a STO (001) substrate.** (**A**) RHEED pattern of a STO (001) substrate. (**B – D**) Instance segmentation masks for three types of features: (B) direct beam (DB), (C) diffraction spots (spot), and (D) diffraction streaks (streak) along with their detection probability (0 – 1).

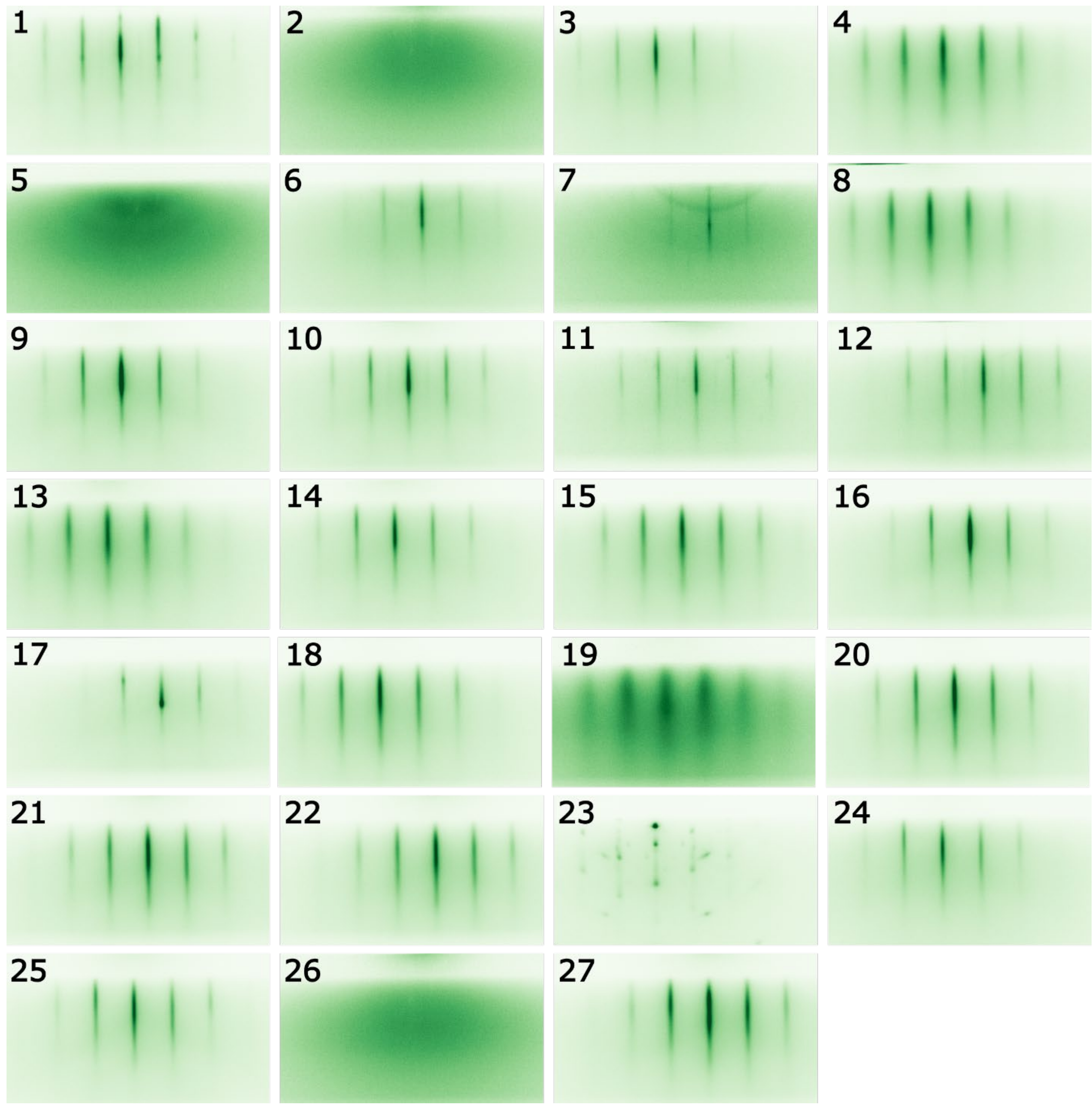

**Fig. S3. RHEED patterns of h-$TbFeO_3$ films at the end of deposition of each autonomous optimization iteration.**

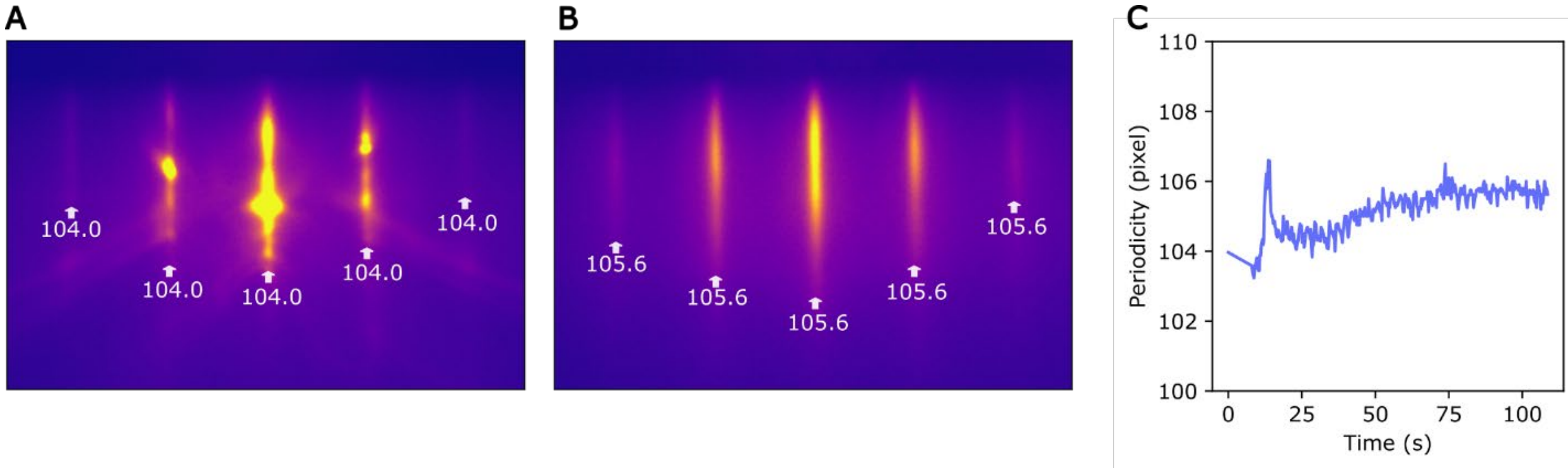

**Fig. S4. Surface structure evolution of h-$TbFeO_3$ deposited under the optimal conditions identified in this work.** (**A**) RHEED pattern of the YSZ (111) substrate captured prior to deposition at the optimal conditions. (**B**) RHEED pattern of h-$TbFeO_3$ captured at the end of the deposition. Periodicities of the YSZ and h-$TbFeO_3$ are labeled in (A) and (B), respectively. (**C**) Periodicity evolution as a function of deposition time. The periodicity increases during growth and ultimately became 1.5 % larger than that of the substrate, indicating the in-plane lattice parameter of h-$TbFeO_3$ relaxed toward a smaller value than the substrate. This implies the film was under in-plane tensile strain.

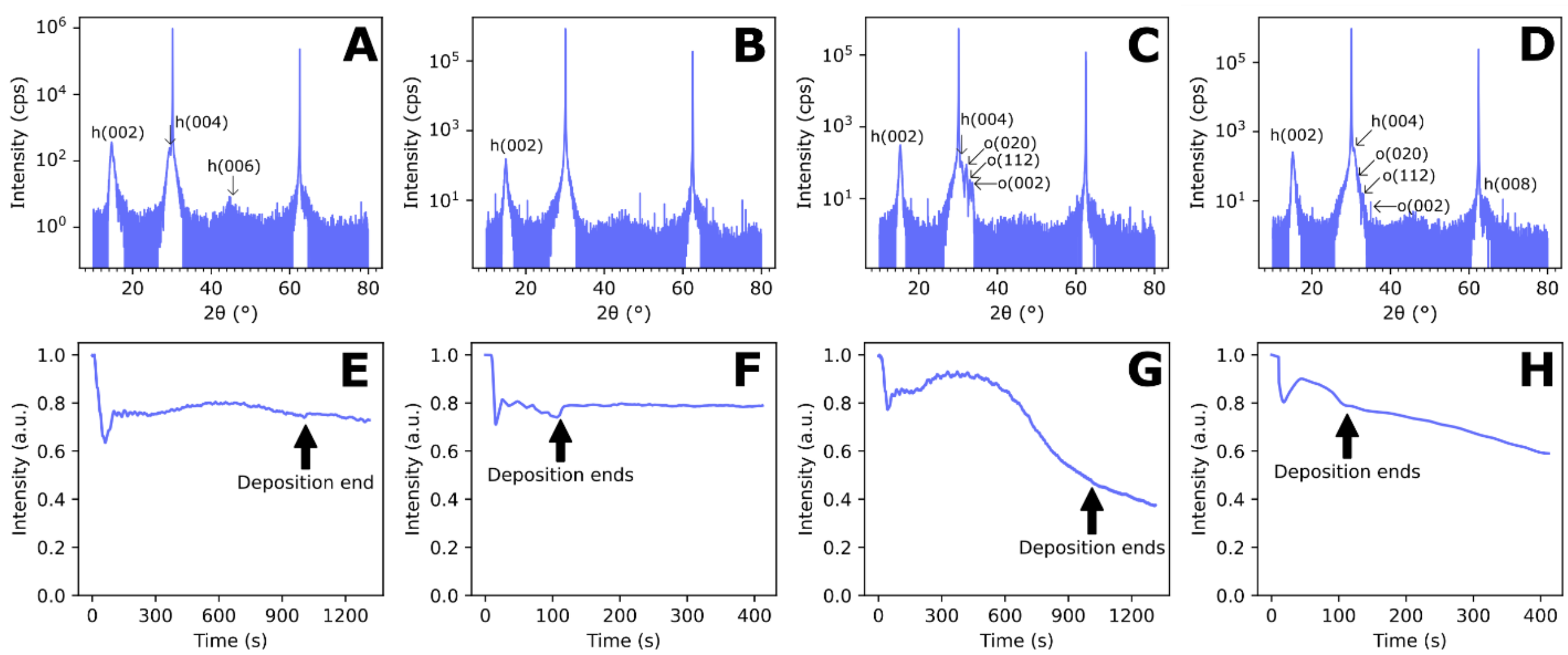

**Fig. S5. Grid measurements of conditions** (a: 1Hz, 831 °C, 1.33 Pa; b: 10 Hz, 831 °C, 1.33 Pa; c: 1 Hz, 1000 °C, 1.33 Pa; d: 10 Hz, 1000 °C, 1.33 Pa) near the optimized conditions. (**A**-**D**) XRD of samples grown at conditions a, b, c and d. In comparison to (A) and (B), (C) and (D) have additional orthorhombic peaks. The h-$TbFeO_3$ and o-$TbFeO_3$ peaks are labeled as "h" and "o", respectively. (**E**-**H**) RHEED intensity vs. time of samples grown at conditions a, b, c, and d.

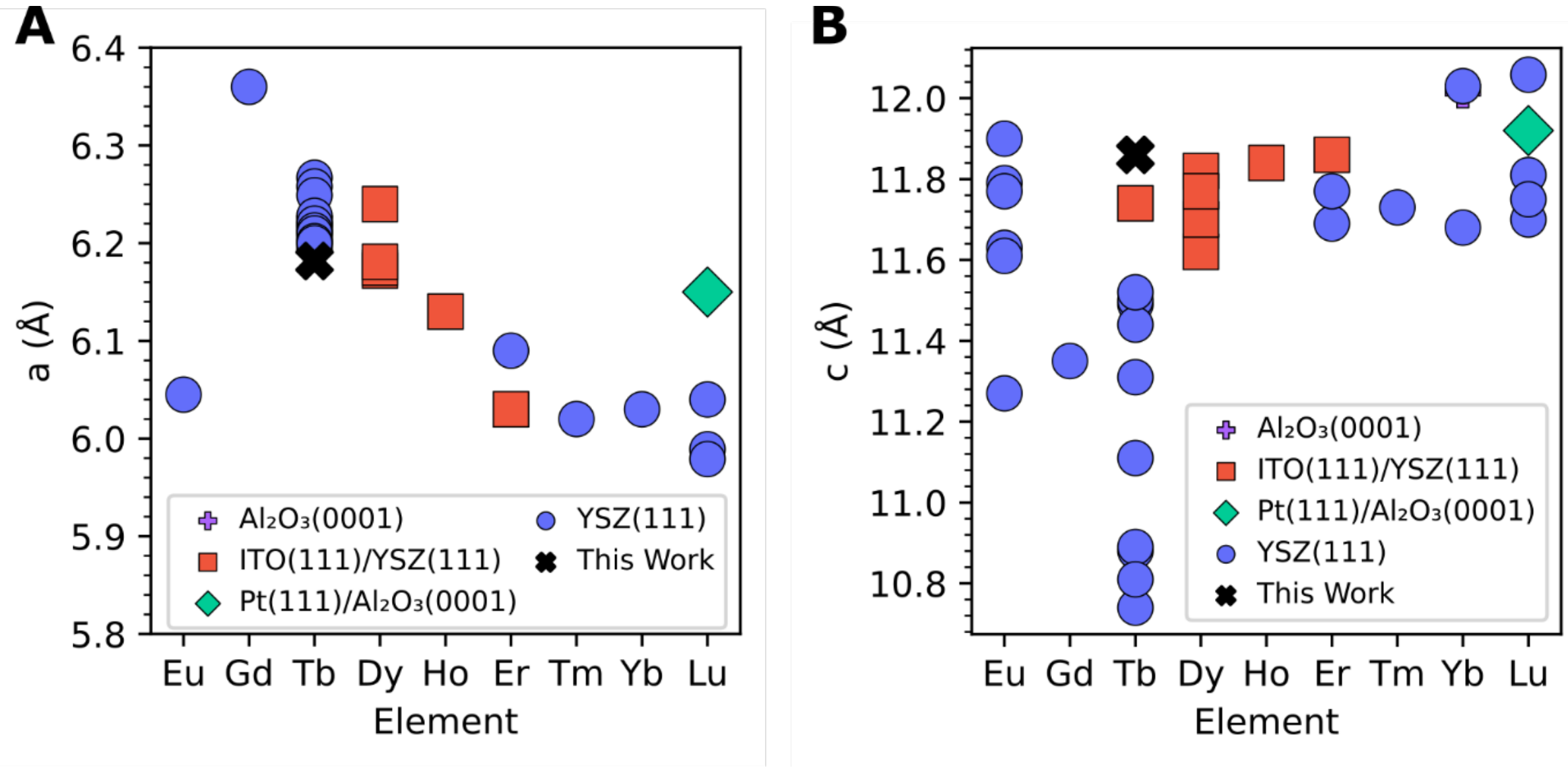

**Fig. S6. Reported growth** (*27*, *37–49*) **of hexagonal (RE)$FeO_3$ process parameters and lattice parameters.** In-plane (A) and out-of-plane (B) lattice parameters of reported (RE)$FeO_3$ structures by its rare earth elements. Lattice parameters of this work are marked with a cross symbol.

**Table. S1. Deposition conditions and the corresponding PM value from the autonomous Run of TbFeO₃** ($w_p = 1$, $w_c = 1$, $w_r = 1$, and $w_s = 0.1$)**.** The top-3 samples with the highest PM values are shown in boldface.

| | **Temperature (°C)** | **Pressure (Pa)** | **Laser Repetition Rate (Hz)** | **PM** |
|---|---|---|---|---|
| 1 | 918 | 4.79E-04 | 9.1 | 1.88 |
| 2 | 400 | 9.37E-01 | 2.0 | 0.49 |
| 3 | 767 | 2.56E-03 | 7.7 | 2.59 |
| 4 | 690 | 5.37E-03 | 2.9 | 2.44 |
| 5 | 518 | 1.18E-03 | 7.1 | 0.59 |
| 6 | 805 | 2.68E-01 | 4.0 | 2.63 |
| 7 | 838 | 8.64E-04 | 2.1 | 1.42 |
| 8 | 714 | 9.28E-01 | 7.7 | 2.43 |
| 9 | 851 | 5.32E-01 | 9.9 | 2.82 |
| 10 | 968 | 1.35E+00 | 6.8 | 2.73 |
| 11 | 1000 | 1.33E+00 | 1.0 | 1.52 |
| 12 | 1000 | 2.80E-01 | 10.0 | 2.16 |
| 13 | 675 | 1.36E+00 | 1.0 | 2.34 |
| 14 | 857 | 1.33E+00 | 6.9 | 2.73 |
| 15 | 786 | 1.40E-02 | 10.0 | 2.57 |
| 16 | 857 | 1.36E-02 | 7.1 | 2.73 |
| 17 | 1000 | 1.43E-02 | 5.5 | 2.55 |
| 18 | 916 | 2.75E-01 | 7.2 | 2.56 |
| 19 | 792 | 2.68E-01 | 7.2 | 2.04 |
| 20 | 890 | 2.73E-01 | 7.8 | 2.68 |
| 21 | 877 | 1.35E+00 | 10.0 | 2.82 |
| **22** | **890** | **1.35E+00** | **10.0** | **2.86** |
| 23 | 734 | 1.47E-04 | 10.0 | 1.50 |
| 24 | 864 | 2.75E-01 | 10.0 | 2.75 |
| **25** | **877** | **1.33E+00** | **10.0** | **2.96** |
| 26 | 532 | 1.36E+00 | 10.0 | 0.82 |
| **27** | **831** | **1.39E+00** | **10.0** | **2.87** |

**Table. S2. Deposition conditions and the corresponding PM value from the autonomous Run of $GdFeO_3$** ($w_p = 1$, $w_c = 1$, $w_r = 1$, and $w_s = 0.1$)**.** The top-3 samples with the highest PM values are highlighted.

| | **Temperature (°C)** | **Pressure (Pa)** | **Laser Repetition Rate (Hz)** | **PM** |
|---|---|---|---|---|
| 1 | 457 | 4.67E-03 | 3.1 | 0.49 |
| 2 | 541 | 2.08E-04 | 5.8 | 2.26 |
| 3 | 970 | 3.49E-01 | 1.5 | 0.37 |
| 4 | 678 | 1.36E-02 | 1.6 | 0.71 |
| 5 | 751 | 1.60E-01 | 8.1 | 2.24 |
| 6 | 350 | 1.99E-04 | 10.0 | 0.63 |
| 7 | 981 | 2.13E-01 | 6.0 | 1.65 |
| 8 | 350 | 1.45E-04 | 1.0 | 0.46 |
| 9 | 1000 | 1.47E-04 | 8.4 | 2.29 |
| 10 | 350 | 1.41E-04 | 6.8 | 0.53 |
| 11 | 799 | 1.51E-04 | 10.0 | 1.39 |
| 12 | 844 | 1.52E-04 | 6.5 | 2.34 |
| 13 | 695 | 1.27E+00 | 4.7 | 2.34 |
| 14 | 1000 | 1.44E-04 | 3.3 | 2.37 |
| 15 | 864 | 1.44E-04 | 4.9 | 0.69 |
| 16 | 1000 | 1.30E+00 | 7.1 | 1.43 |
| **17** | **883** | **1.27E+00** | **6.8** | **2.57** |
| **18** | **870** | **1.31E+00** | **6.8** | **2.55** |
| **19** | **851** | **1.31E+00** | **6.8** | **2.53** |

**Table. S3. Deposition conditions and the corresponding PM value from the autonomous Run of $EuFeO_3$** ($w_p = 1$, $w_c = 1$, $w_r = 0$, and $w_s = 1$). The top-3 samples with the highest PM values are highlighted.

| | Temperature (°C) | Pressure (Pa) | Laser Repetition Rate (Hz) | Laser Power (mJ) | PM |
|---|---|---|---|---|---|
| 1 | 500 | 2.47E-04 | 4.0 | 136 | 0.47 |
| 2 | 850 | 1.59E-03 | 4.0 | 136 | 1.22 |
| 3 | 900 | 1.75E-03 | 10.0 | 136 | 0.76 |
| 4 | 700 | 1.75E-03 | 2.0 | 136 | 0.16 |
| 5 | 840 | 1.83E-03 | 3.0 | 146 | 0.91 |
| 6 | 850 | 9.25E-04 | 3.5 | 125 | 2.02 |
| 7 | 840 | 9.91E-04 | 20.0 | 125 | 2.27 |
| **8** | **810** | **1.40E-03** | **20.0** | **125** | **2.78** |
| 9 | 580 | 4.87E-02 | 20.0 | 125 | 1.03 |
| 10 | 750 | 1.64E-03 | 20.0 | 125 | 2.69 |
| **11** | **840** | **1.92E-03** | **20.0** | **146** | **2.73** |
| 12 | 810 | 1.81E-03 | 18.0 | 146 | 2.59 |
| **13** | **1000** | **4.53E-02** | **20.0** | **146** | **2.73** |

**Table. S4. Reported values of the lattice parameters of $REFeO_3$.**

| Materials | Refs | Method | Substrate | a (Å) | c (Å) |
|---|---|---|---|---|---|
| $TbFeO_3$ | Magnetic Phase Transition-Induced Modulation of Ferroelectric Properties in Hexagonal RFeO3 (R = Tb and Ho)(*37*) | PLD | ITO (111)/YSZ (111) | 6.21 | 11.74 |
| $HoFeO_3$ | Magnetic Phase Transition-Induced Modulation of Ferroelectric Properties in Hexagonal RFeO3 (R = Tb and Ho)(*37*) | PLD | ITO (111)/YSZ (111) | 6.13 | 11.84 |
| $TbFeO_3$ | Emergence of critical thickness and multifaceted role of stacking faults in high misfit hexagonal ferrites films(*38*) | PLD | YSZ (111) | 6.267 | 10.88 |
| $TbFeO_3$ | Emergence of critical thickness and multifaceted role of stacking faults in high misfit hexagonal ferrites films(*38*) | PLD | YSZ (111) | 6.258 | 10.89 |
| $TbFeO_3$ | Emergence of critical thickness and multifaceted role of stacking faults in high misfit hexagonal ferrites films(*38*) | PLD | YSZ (111) | 6.249 | 10.74 |
| $TbFeO_3$ | Emergence of critical thickness and multifaceted role of stacking faults in high misfit hexagonal ferrites films(*38*) | PLD | YSZ (111) | 6.228 | 10.81 |
| $TbFeO_3$ | Emergence of critical thickness and multifaceted role of stacking faults in high misfit hexagonal ferrites films(*38*) | PLD | YSZ (111) | 6.219 | 11.11 |
| $TbFeO_3$ | Emergence of critical thickness and multifaceted role of stacking faults in high misfit hexagonal ferrites films(*38*) | PLD | YSZ (111) | 6.22 | 11.31 |
| $TbFeO_3$ | Emergence of critical thickness and multifaceted role of stacking faults in high misfit hexagonal ferrites films(*38*) | PLD | YSZ (111) | 6.213 | 11.49 |
| $TbFeO_3$ | Emergence of critical thickness and multifaceted role of stacking faults in high misfit hexagonal ferrites films(*38*) | PLD | YSZ (111) | 6.21 | 11.5 |
| $TbFeO_3$ | Emergence of critical thickness and multifaceted role of stacking faults in high misfit hexagonal ferrites films(*38*) | PLD | YSZ (111) | 6.202 | 11.44 |
| $TbFeO_3$ | Emergence of critical thickness and multifaceted role of stacking faults in high misfit hexagonal ferrites films(*38*) | PLD | YSZ (111) | 6.2 | 11.52 |

| $GdFeO_3$ | Epitaxial growth of hexagonal GdFeO3 thin films with magnetic order by pulsed laser deposition(*39*) | PLD | YSZ (111) | 6.36 | 11.35 |
|---|---|---|---|---|---|
| $LuFeO_3$ | Magnetic Structure and Ordering of Multiferroic Hexagonal LuFeO3(*40*) | MBE | YSZ (111) | 5.989 | 11.7 |
| $LuFeO_3$ | Magnetic Structure and Ordering of Multiferroic Hexagonal LuFeO3(*40*) | MBE | YSZ (111) | 5.979 | 11.81 |
| $EuFeO_3$ | XRD and HREM Studies of Epitaxially Stabilized Hexagonal Orthoferrites RFeO3 (R = Eu−Lu)(*41*) | MOCVD | YSZ (111) | | 11.27 |
| $ErFeO_3$ | XRD and HREM Studies of Epitaxially Stabilized Hexagonal Orthoferrites RFeO3 (R = Eu−Lu)(*41*) | MOCVD | YSZ (111) | 6.09 | 11.69 |
| $TmFeO_3$ | XRD and HREM Studies of Epitaxially Stabilized Hexagonal Orthoferrites RFeO3 (R = Eu−Lu)(*41*) | MOCVD | YSZ (111) | 6.02 | 11.73 |
| $YbFeO_3$ | XRD and HREM Studies of Epitaxially Stabilized Hexagonal Orthoferrites RFeO3 (R = Eu−Lu)(*41*) | MOCVD | YSZ (111) | 6.03 | 11.68 |
| $LuFeO_3$ | XRD and HREM Studies of Epitaxially Stabilized Hexagonal Orthoferrites RFeO3 (R = Eu−Lu)(*41*) | MOCVD | YSZ (111) | 6.04 | 11.75 |
| $LuFeO_3$ | Structure Quality of LuFeO3 Epitaxial Layers Grown by Pulsed-Laser Deposition on Sapphire/Pt(*56*) | PLD | Pt (111)/$Al_2O_3$ (0001) | 6.15 | 11.92 |
| $DyFeO_3$ | Room-Temperature Antiferroelectricity in Multiferroic Hexagonal Rare-Earth Ferrites(*57*) | PLD | ITO (111)/YSZ (111) | 6.24 | 11.82 |
| $EuFeO_3$ | Examination of ferroelectric and magnetic properties of hexagonal ErFeO3 thin films(*58*) | PLD | YSZ (111) | | 11.9 |
| $EuFeO_3$ | Examination of ferroelectric and magnetic properties of hexagonal ErFeO3 thin films(*58*) | PLD | YSZ (111) | | 11.79 |
| $EuFeO_3$ | Examination of ferroelectric and magnetic properties of hexagonal ErFeO3 thin films(*58*) | PLD | YSZ (111) | 6.045 | 11.77 |
| $EuFeO_3$ | Examination of ferroelectric and magnetic properties of hexagonal ErFeO3 thin films(*58*) | PLD | YSZ (111) | | 11.63 |
| $EuFeO_3$ | Examination of ferroelectric and magnetic properties of hexagonal ErFeO3 thin films(*58*) | PLD | YSZ (111) | | 11.61 |
| $YbFeO_3$ | Ferroelectricity and Ferrimagnetism of Hexagonal YbFeO3 Thin Films(*42*) | PLD | YSZ (111) | | 11.76 |
| $ErFeO_3$ | Ferroelectricity and weak ferromagnetism of hexagonal ErFeO3 thin films(*43*) | PLD | YSZ (111) | | 11.77 |
| $ErFeO_3$ | Ferroelectric and Magnetic Properties of Hexagonal ErFeO3 Epitaxial Films(*50*) | PLD | ITO (111)/YSZ (111) | 6.03 | 11.86 |
| $LuFeO_3$ | Epitaxial growth and magnetic properties of h-LuFeO3 thin films(*45*) | PLD | YSZ (111) | | 12.0582 |
| $DyFeO_3$ | Antiferroelectric-to-ferroelectric phase transition in hexagonal rare-earth iron oxides(*47*) | PLD | ITO (111)/YSZ (111) | 6.172 | 11.62 |
| $DyFeO_3$ | Antiferroelectric-to-ferroelectric phase transition in hexagonal rare-earth iron oxides(*47*) | PLD | ITO (111)/YSZ (111) | 6.175 | 11.7 |
| $DyFeO_3$ | Antiferroelectric-to-ferroelectric phase transition in hexagonal rare-earth iron oxides(*47*) | PLD | ITO (111)/YSZ (111) | 6.181 | 11.77 |
| $YbFeO_3$ | Structural studies and physical properties of hexagonal-YbFeO3 thin films(*49*) | radio-frequency (RF) magnetron sputtering | $Al_2O_3$ (0001) | | 12.02 |
| $YbFeO_3$ | Structural studies and physical properties of hexagonal-YbFeO3 thin films(*49*) | radio-frequency (RF) magnetron sputtering | YSZ (111) | | 12.03 |

**Video. S1. A demonstration of a RHEED recording processed through the RHEED image analysis pipeline is available at the link:** https://umd.box.com/s/xtv708n7fjibszf5ockgixg60tcsh3t8**.**